\documentclass[sigconf]{acmart}

\usepackage{booktabs} 
\usepackage{todonotes}
\usepackage{changebar}
\usepackage{soul}
\usepackage{color}

\setcopyright{rightsretained}

\acmDOI{}

\acmISBN{}

\acmConference{SysML Conference}
\acmYear{2018}
\copyrightyear{2018}

\acmArticle{}
\acmPrice{}

\begin{document}

\newcommand{\othertm}{\textsuperscript{$\star$}}
\newcommand{\regtm}{\textsuperscript{\textregistered{}}}
\newcommand{\tm}{{\scriptsize\texttrademark{}}}
\newcommand{\CPUtm}{Intel\textsuperscript{\textregistered} Xeon\textsuperscript{\textregistered}~processor\xspace}
\newcommand{\partCPUtm}{Intel\textsuperscript{\textregistered} Xeon\textsuperscript{\textregistered}\xspace}
\newcommand{\partMICtm}{Intel\textsuperscript{\textregistered} Xeon Phi\textsuperscript{\texttrademark}\xspace}
\newcommand{\CPU}{Intel Xeon~processor\xspace}
\newcommand{\partMIC}{Intel Xeon Phi\xspace}
\newcommand{\partCPU}{Intel Xeon\xspace}
\newcommand{\Intel}{Intel\textsuperscript{\textregistered}\xspace}
\newcommand{\VTunetm}{VTune\textsuperscript{\textregistered}~Amplifier\xspace}
\newcommand{\mathtext}[1]{\text{\textit{#1}}}


\title{On Scale-out Deep Learning Training for Cloud and HPC}

\author{Srinivas Sridharan, Karthikeyan Vaidyanathan, Dhiraj Kalamkar, Dipankar Das,\\Mikhail E. Smorkalov, Mikhail Shiryaev, Dheevatsa Mudigere, Naveen Mellempudi,\\Sasikanth Avancha, Bharat Kaul, Pradeep Dubey}

\affiliation{%
  \institution{Intel Corporation}
  \streetaddress{}
  \city{} 
  \state{} 
  \postcode{}
}
\email{[srinivas.sridharan, karthikeyan.vaidyanathan, dhiraj.kalamkar, dipankar.das, mikhail.e.smorkalov, mikhail.shiryaev, dheevatsa.mudigere, naveen.mellempudi, sasikanth.avancha, bharat.kaul, pradeep.dubey]@intel.com}







\renewcommand{\shortauthors}{Srinivas Sridharan, Karthikeyan Vaidyanathan, Dhiraj Kalamkar, Dipankar Das, Mikhail E. Smorkalov, Mikhail Shiryaev,\\Dheevatsa Mudigere, Naveen Mellempudi,Sasikanth Avancha, Bharat Kaul, Pradeep Dubey}





\maketitle

 \section*{Introduction}\label{sec:intro}
Deep Learning (DL) is driving the adoption of Machine Learning (ML) and Artificial Intelligence (AI) across a wide range of application domains such as image recognition, natural language processing, and autonomous driving. The exponential growth in use of large deep neural networks has accelerated the need for training these deep neural networks in hours or even minutes. This can only be achieved through
scalable and efficient distributed training, since a single node/card cannot satisfy the compute, memory, and I/O requirements of today's state-of-the-art deep neural networks. However, scaling synchronous \emph{Stochastic Gradient Descent} (SGD) is still a challenging problem and requires continued research/development. This entails innovations spanning algorithms \cite{DBLP:journals/corr/KeskarMNST16,DBLP:journals/corr/GoyalDGNWKTJH17}, frameworks \cite{intelcaffe,tensorflow,ngraph}, communication libraries \cite{mlsl,DBLP:journals/corr/abs-1708-02188,nccl}, and system design \cite{Jouppi:2017:IPA:3079856.3080246}. In this paper, we describe the philosophy, design, and implementation of Intel\textregistered \emph{Machine Learning Scaling Library} (MLSL) and present proof-points demonstrating DL training on 100s to 1000s of nodes across Cloud and HPC systems. 


\section*{Design Choices and Insights}\label{sec:theory}
The common parallelization techniques for partitioning work across multiple nodes, are \emph{data parallelism}~(replicating the entire model) and \emph{model parallelism}~(distributing the model). In \cite{DBLP:journals/corr/0002AMVSKKD16}, we present a detailed theoretical analysis of computation and communication involved in DL training. Based on this analysis, we derived the  \emph{compute to communication ratio} that captures the number of compute operations per layer to the communication volume. The goal is to maximize this ratio for best scaling. For data parallelism, we observe that this ratio is a function of the size of output feature-maps, mini-batch size and effectiveness of overlap. Interestingly, it does not depend on the kernel size or number of input/output feature maps or stride. We use these insights to guide different design choices for realizing scalable distributed training.

\noindent\textbf{\emph{Choosing the right work partitioning strategy:}} First, using the methodology in \cite{DBLP:journals/corr/0002AMVSKKD16}, we identify the optimal parallelization strategy for each layer depending on the type of the layer (convolutional, fully connected, etc.), size of output feature maps, and so on. Further, we developed a novel work partition strategy, called \emph{hybrid parallelism}, which partitions the work across both the data and model using the concept of \emph{node groups}; i.e. nodes within a group employ model parallelism and data parallelism is used across groups. One could consider data and model parallelism as two extreme design points of hybrid parallelism with node group size being one and all nodes respectively. 
\noindent\textbf{\emph{Increasing concurrency with large batch training:}} 
For data parallelism, we observe that the compute to communication ratio is proportional to the mini-batch size. This implies, scaling will be negatively impacted as we strong-scale the mini-batch and the mini-batch per node drops. More specifically, communication starts dominating total execution time since communication tends to become more latency bound and there's lesser compute to hide communication. Hence, large batch training is essential for efficient scaling and this observation is in line with recent efforts enabling large mini-batch in training without affecting accuracy \cite{DBLP:journals/corr/KeskarMNST16,DBLP:journals/corr/GoyalDGNWKTJH17,DBLP:journals/corr/abs-1711-00489}. 

\noindent\textbf{\emph{Overlapping communication and computation:}}
Unlike model parallelism, in data parallelism, there's significant opportunity to overlap communication with compute. Each node computes partial weight gradients for its mini-batch in the back-propagation step in each layer and aggregates these partial gradients across all nodes using an \emph{allreduce} operation. These aggregated weight gradients are used to update the weights and only required right before the forward propagation step for that layer in the next iteration.
This is captured in the compute to communication ratio and relies on networking library/HW's ability to \emph{asynchronously progress} communication and framework's ability to schedule communication to maximize compute-communication overlap. 

\noindent\textbf{\emph{Prioritizing latency-bound communication:}}
While overlapping communication with computation across
layers is indispensable, the overheads of the first layer's weight gradient communication in data parallelism is fully exposed given lack of useful compute to overlap communication. 
In other words, while network bandwidth is critical for all other layers, optimizing for network latency is essential for the first layer since size of the weight gradients are typically small(er). 
This motivates the need for further prioritizing and completing the first layers communication operations before communication operations from later layers even though they were issued earlier. Similarly, in the case of model/hybrid parallelism, activation communication must be prioritized as they may block the next layer's compute.

\noindent\textbf{\emph{Reducing communication volume:}}
Finally, scaling can be further improved by reducing the volume of communicated data. For instance, this can be achieved through message compression and/or quantization \cite{seide20141,dettmers20158,lin2017deep}. The growing adoption of lower precision for training, has an impact of communication/scaling as well. At a minimum, while communication should support the same precision as the compute, the precision for communication could be further reduced allowing for improved scaling. However, this entails frameworks, libraries and HW to natively support low precision communication, for guaranteeing correctness and realizing the performance potential. 


We now present Intel\textregistered \emph{Machine Learning Scaling Library} (MLSL), a core component of our solution stack embodying many of the optimizations and design choices just described.

\section*{Machine Learning Scalability Library}\label{sec:mlsl}

\begin{figure}[h]
\includegraphics[height=1.25in, width=2.5in]{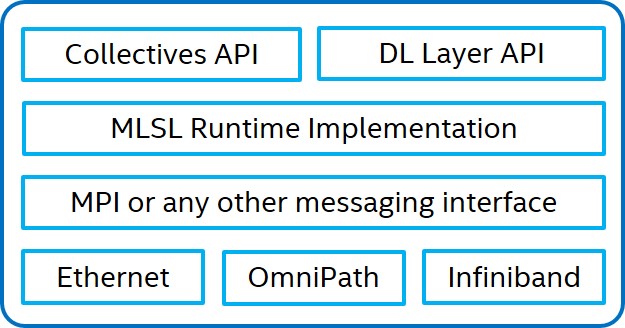}
\caption{Intel MLSL Architecture}
\label{fig:mlsl-arch}
\end{figure}
\vspace{-2mm}

Figure \ref{fig:mlsl-arch} presents the MLSL SW architecture. At the highest level, MLSL exposes two interfaces for frameworks: \emph{collectives} and \emph{DL layer}. The collectives API is similar to \emph{Message Passing Interface} (MPI) collectives interface and supports commonly used collective operations found in DL/ML workloads. The DL Layer API is a higher-level interface that abstracts the exact communication operation depending on the type of parallelism chosen (data, model, or hybrid) for each layer of the neural network at runtime, thus reducing the hassle of supporting these different scenarios within each framework explicitly. 


Regardless of the chosen interface, MLSL's runtime implementation enables novel DL specific optimizations unavailable in MPI and other communication libraries, such as asynchronous progress for compute-communication overlap, dedicating one or more cores for driving the network in an optimal manner, message prioritization, and collectives in low-precision data types. MLSL's flexible API enables these runtime optimizations to be applied across frameworks and lowers the effort required in optimizing each framework independently. Furthermore, MLSL uses existing communication libraries, such as MPI, for commonly used control path operations but only implements performance critical data path operations in an optimal manner.

The benefits of MLSL's design and implementation becomes self-evident when examining one of the DL-specific communication optimizations in greater detail. Like mentioned earlier, with data parallelism the weight gradient communication in the first layer is latency bound and the updated weights are required immediately in the forward pass. However, MPI interface and implementations do not support prioritizing such messages. MLSL's \emph{message prioritization} feature overcomes this limitation by preempting an ongoing large weight gradient exchange operation from one of the later layers and instead prioritizes the smaller weight gradient allreduce from the first layer to proceed. The preempted operations are completed in an optimal manner as and when they are required in the forward pass and not necessarily the order in which they were originally issued. This optimization resulted in 1.8x to 2.2x reduction in exposed communication time for standard topologies such as Resnet-50, VGG-16, and Googlenet on Intel\textsuperscript{\textregistered} Xeon\textsuperscript{\textregistered}~Gold 6148 processors (code-named Skylake) and 10Gbps Ethernet. Additional DL specific optimizations, such as message quantization and persistent collectives \cite{Morgan:2017:PPP:3127024.3127028}, are currently being evaluated and will be made available as part of upcoming MLSL SW releases. 



\begin{figure}[h]
\includegraphics[height=1.8in, width=3.5in]{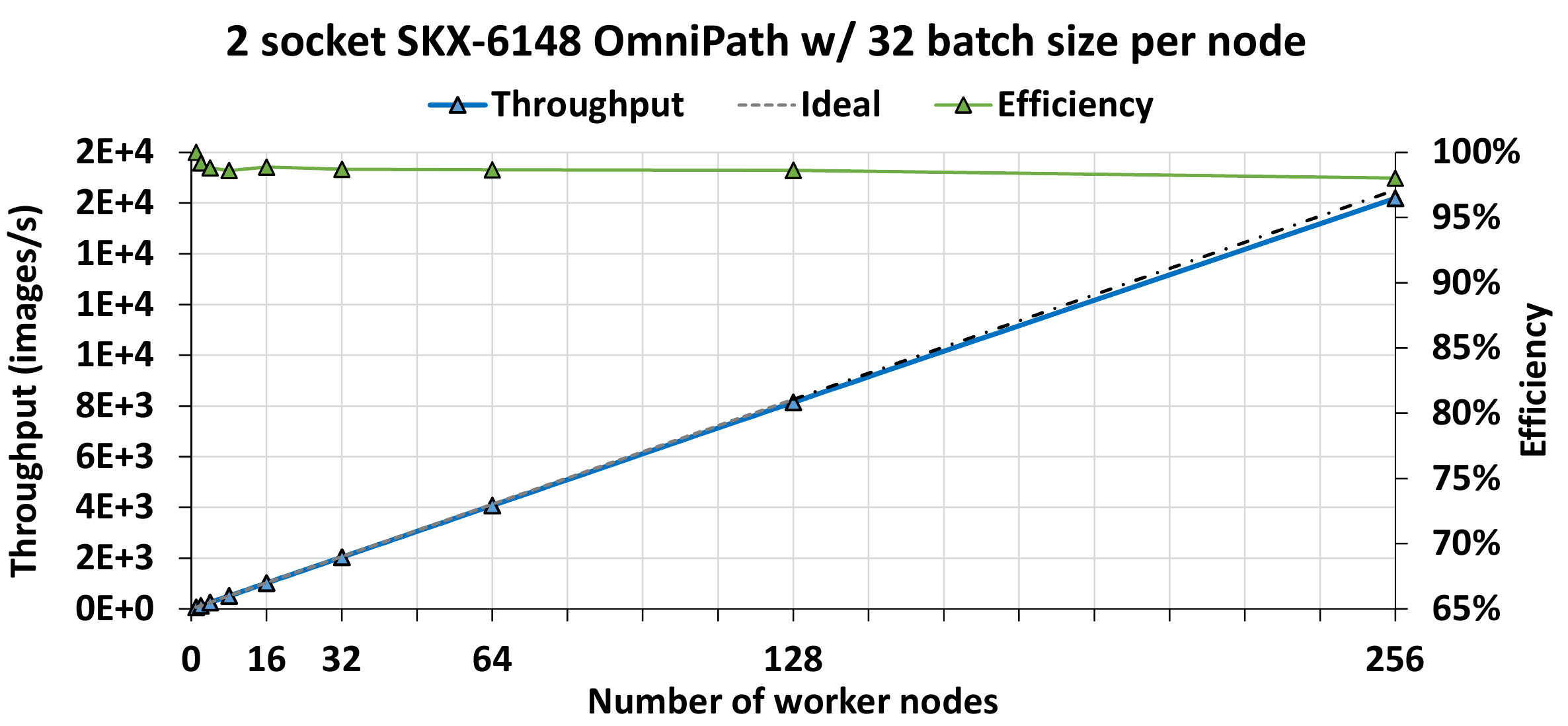}
\caption{Resnet-50 scaling on Intel Xeon / Omnipath}
\label{fig:mlsl-results}
\end{figure}
\vspace{-2mm}

MLSL has been integrated with numerous DL frameworks, including but not limited to, BVLC Caffe \cite{intelcaffe}, Google\othertm TensorFlow, and Intel\textregistered ~nGraph\tm \cite{ngraph}. While the integration strategy differs in each case, the use of a single library facilitates common set of optimizations across all these frameworks. For instance, Figure \ref{fig:mlsl-results} presents Resnet-50 scaling on Intel\textsuperscript{\textregistered} Xeon\textsuperscript{\textregistered}~Gold 6148 processors (code-named Skylake) and Intel Omnipath fabric using Intel Caffe and MLSL demonstrate ~$90\%$ scaling on 256 nodes (75.8\% top-1 validation accuracy). Further, this solution has been used to scale deep neural networks  solving scientific pattern classification problems to ~$9600$ Xeon-Phi nodes \cite{Kurth:2017:DLS:3126908.3126916} and to train Resnet-50 in $40$ minutes on $256$ nodes on the MareNostrum system at Barcelona Supercomputing Center \cite{surfara}. For TF, we have developed a new distributed solution that adopts Uber Horovod \cite{horovod} interface but uses MLSL to achieve higher scaling performance over the out-of-box Horovod MPI implementation. We observe >93\% scaling efficiency on the fore-mentioned Intel\textsuperscript{\textregistered} Xeon\textsuperscript{\textregistered} system on 64 nodes. For nGraph, we added new graph passes to insert non-blocking MLSL collective operations and novel scheduling optimizations to ensure maximum compute-communication overlap. More details on MLSL with TF and nGraph will be shared in the near future. 

We plan to continue extending MLSL with novel DL features and optimizations. We are actively looking for collaborating with researchers/developers interested in using MLSL and extending the scaling envelope for DL workloads.



\bibliographystyle{ACM-Reference-Format}
\bibliography{references} 

\scriptsize{
\noindent\othertm{}Other names and brands may be claimed as property of others.

\noindent
Intel and Xeon are trademarks of Intel Corporation in the U.S. and/or other countries.
Software and workloads used in performance tests may have been optimized
for performance only on Intel microprocessors.  Performance tests, such as
SYSmark and MobileMark, are measured using specific computer systems,
components, software, operations and functions.  Any change to any of those
factors may cause the results to vary.  You should consult other information
and performance tests to assist you in fully evaluating your contemplated
purchases, including the performance of that product when combined with
other products.  For more information go to \url{http://www.intel.com/performance}.
}

\end{document}